%% file: toribio.tex
\documentclass[12pt,preprint]{aastex}

\usepackage{natbib}

\newcommand{\noprint}[1]{}

\newcommand{\smallhi}{{H$\,$\tiny I}}
\newcommand{\hi}{{H$\,$\footnotesize I}}
\newcommand{\kms}{\ km\ s$^{-1}$}
\newcommand{\df}{\mbox{DEF}}

\newcommand{\sdef}{\sigma_\mathrm{DEF}}
\newcommand{\wtwenty}{W_\mathrm{20}}
\newcommand{\wfifty}{W_\mathrm{50}}

\newcommand{\subhi}{_\mathrm{H\mbox{\tiny I}}}

\newcommand{\nhi}{N\subhi}
\newcommand{\shi}{\overline{\Sigma}\subhi}

\shorttitle{\hi\ Distribution and Distances of Gas-Poor Spirals}
\shortauthors{Toribio \& Solanes}

\begin{document}

\title{\hi\ Distribution and Tully-Fisher Distances of Gas-Poor Spiral Galaxies
in the Virgo Cluster Region}


\author{M.\ Carmen Toribio and Jos\'e M.\ Solanes}
\affil{Departament d'Astronomia i Meteorologia and Institut de Ci\`encies del Cosmos, Universitat de Barcelona. C/ Mart\'{\i} i Franqu\`es, 1; E--08028~Barcelona, Spain}
\email{mctoribio@am.ub.es, jm.solanes@ub.edu}

\begin{abstract}

  We present aperture synthesis observations in the 21 cm line of
  pointings centered on the Virgo Cluster region spirals NGC 4307, NGC
  4356, NGC 4411B, and NGC 4492 using the Very Large Array (VLA) 
  radiotelescope in its CS configuration. These galaxies were
  identified in a previous study of the three-dimensional distribution
  of \hi\ emission in the Virgo region as objects with a substantial
  dearth of atomic gas and Tully-Fisher (TF) distance estimates that
  located them well outside the main body of the cluster. We have
  detected two other galaxies located in two of our fields and
  observed bands, the spiral NGC 4411A and the dwarf spiral VCC
  740. We provide detailed information of the gas morphology and
  kinematics for all these galaxies. Our new data confirm the
  strong \hi\ deficiency of all the main targets but NGC 4411B, which
  is found to have a fairly normal neutral gas content. The VLA
  observations have also been used to discuss the applicability of TF
  techniques to the five largest spirals we have observed. We conclude
  that none of them is actually suitable for a TF distance evaluation,
  whether due to the radical trimming of their neutral hydrogen disks
  (NGC 4307, NGC 4356, and NGC 4492) or to their nearly face-on
  orientation (NGC 4411A and B). 

\end{abstract}


\section{Introduction}

The regions around rich clusters are the most obvious sites to
evidence the transformation of galaxy disks driven by the surrounding
intracluster medium (ICM). The increased density of both hot gas and
galaxies, as well as the high relative velocities of the latter, set
the scene for dramatic effects on their fragile interstellar medium
(ISM).

In the local universe, the nearby Virgo Cluster region is an ideal
place to quantify these nurturing effects, because its proximity makes
it possible to probe the gaseous disks with higher sensitivity and
resolution than in any other cluster. Another characteristic that
makes this galaxy system very appealing for studies of galaxy
evolution is its relative dynamical youth: Virgo has a central region
with several substructures in the process of merging, surrounded by
suburbia dominated by late-type galaxies that might fall into the
cluster during the next Hubble time. As noted by \citet*{Vol01} and
\citet*{Sol01}, environmental mechanisms such as ram-pressure
stripping may see their effectiveness increased during the built-up of
clusters.

There is a long list of studies covering a broad stretch of the
electromagnetic spectrum that have investigated the impact of cluster
residency on the late-type galaxy population. Plenty of them use data
from the 21-cm emission line of the abundant, and easy to strip off,
neutral hydrogen (\hi) of the disks, as the most direct approach to
measure the affectation of the ISM. These investigations generally
agree in indicating that Virgo spirals, like those inhabiting other
rich clusters, tend to have less neutral gas than their field
counterparts, and also in finding evidence for a correlation of the
\hi\ deficiency with clustercentric distance, with \hi-poor disks
typically situated close to the cluster cores and galaxies removed
from those regions showing normal gas contents
\citep[e.g.][]{HG86,Cay94,Sol01,Gav05,Chu08}. The lack of atomic gas,
which usually affects the outer disks, is frequently accompanied by an
even more severe truncation of the H$\alpha$ emission and the
corresponding quenching of star formation beyond that truncation
radius \citep[e.g.][]{KK04,CK08}. Recently, there are also evidences
than it could be associated with H$_{\mathrm{2}}$ reduction too
\citep{Fum09}.

Virgo is also the first, and so far the only, cluster region for which
the \emph{spatial} distribution of the \hi\ deficiency has been mapped
\citep*[hereafter Sol02]{Sol02}. By using homogenized Tully-Fisher
(TF) distance moduli and 21 cm data from single-dish observations for
161 galaxies, these authors confirmed that the neutral gas deficiency
in the Virgo Cluster decreases with increasing 3D barycentric
distance. This study, however, also revealed the presence of an
unexpectedly large fraction of strongly \hi-deficient spirals with TF
radial distances pointing to a location well outside the cluster body.

Ensuing studies based on both analytic infall models \citep*{San02} and
$N$-body simulations \citep*{Mam04} investigated the possibility that
some of the gas-poor spirals in Virgo's suburbia had lost their gas
content in a previous passage through the cluster core and were now
lying near the apocenter of their orbits. Both works lead to
the identification by \citet*{San04} of 13 extremely \hi-poor spiral
galaxies for which the lack of cold neutral gas could hardly be
attributed to ISM-ICM stripping, unless their radial distances
were affected by relative errors much larger than the typical
uncertainty attributed to TF measurements. Other possibilities for the
origin of these \hi\ outliers, such as gas deficiency caused by
gravitational interactions (tides or mergers) with companion galaxies, 
or errors in the \hi\ deficiency estimates arising from morphological
misclassifications, were also investigated and considered less
probable.

With the aim of shedding more light on this matter, we initiated some
time ago a program of dedicated observations of some of the outlying
\hi-deficient Virgo Cluster spirals found in \cite{San04}. In this
paper, we attempt to improve the results of the aforementioned study,
which were based on the analysis of integral galaxy properties
retrieved from public databases, by investigating the neutral gas
distribution and kinematics, as well as the TF distances, for 4 of
these objects by means of deep 21 cm synthesis observations carried
out with the Very Large Array (VLA) in its CS
configuration\footnote{The VLA is a facility of the National Radio
  Astronomy Observatory.}. The paper begins by describing in
Section~\ref{selection} the selection of the targets. The acquisition
and reduction of the 21-cm line data, as well as the steps
followed in the derivation of the \hi\ synthesis results are discussed
in Section~\ref{processing}. In Section~\ref{analysis}, we analyze
case by case the \hi\ properties of all the galaxies showing 21 cm
emission in the selected fields of view, while in Section~\ref{tf} we
discuss the applicability of the TF technique to the five large
spirals that have been observed. Finally, the results and
conclusions of this work are given in Section~\ref{conclusions}.

\section{Galaxy Selection}\label{selection}

We have used the VLA to observe Virgo Cluster galaxies that are faint
in the 21 cm line. Three of the targets, NGC 4307, NGC 4356, and NGC
4492, belong to the subset of 13 \hi-outliers identified by
\citet{San04}\footnote{Seven other members of this subset are among
  the targets of the VIVA (VLA Imaging of Virgo in Atomic gas) survey
  by J.\ Kenney, J.\ van Gorkom, and cols.\ in which the gas is imaged
  down to a column-density sensitivity of a few times $10^{19}$\
  cm$^{-2}$, similar to that of the present observations.}. These are
galaxies with neutral gas deficiencies deviating by more than $3\sdef$
from normalcy. A fourth pointing has been centered on NGC 4411B,
another spiral with a less extreme \hi\ deficiency. All these objects
are among the most gas-poor spiral galaxies lying on the sky between
the M49 subcluster and the W'/W cloud region (see Fig.~\ref{location})
and have TF estimates of their radial distances suggestive of a
possible background location far from the cluster core, provided one
adopts the currently preferred Virgo mean distance of
$d_{\mathrm{Virgo}}\sim 16-17$ Mpc, as suggested by both the
measurements of $H_0$ from HST observations of Cepheids and the
spatial distribution of the early type galaxy population and X-ray gas 
\citep*[e.g.][]{Gav99,Fre02,San04,Mei07}. Another characteristic
these galaxies have in common is that their systemic velocities do not
differ much from the mean cluster velocity. Galaxy properties are
compiled in Table~\ref{toribio_tab1}, including a preliminary estimate
of their neutral gas deficiency using the following
distance-independent calibrator
\begin{equation}\label{def} 
\df=\langle\log\shi (T)\rangle-\log\shi\;,
\end{equation}
which compares the logarithms of the expected and observed values of
the \emph{hybrid} \hi\ surface density calculated from the ratio
between the intrinsic integrated \hi\ flux and the square of the
apparent major optical diameter of a galaxy of morphological type
$T$. We have followed \citeauthor{Sol02} and adopted for $\langle\log\shi
(T)\rangle$ the values: 0.24 for Sa, Sab types; 0.38 for Sb; 0.40 for
Sbc; 0.34 for Sc; and 0.42 for later types in units of Jy\kms\ per
arcmin square. Thus, taking into account that the rms scatter in $\df$
for field galaxies is $\sdef =0.24$, an object with $\df > 3\sdef$ has
less than $20\%$ of the expected \hi\ mass for a galaxy of its
morphology.

We have also included in Table~\ref{toribio_tab1} four more galaxies
located in some of the fields of view of our main targets. These are:
VCC 740, a small spiral in the vicinity of NGC 4356; the first
component of the pair NGC 4411A/B, very close to its companion galaxy
in both projected position and radial velocity (their centers are
separated only by $4\arcmin$ and 11\kms, respectively), but whose
estimated TF radial distance of $\sim 16$ Mpc (\citeauthor{Sol02}) indicates
that it is likely a Virgo Cluster member; and two dwarf ellipticals,
VCC numbers 933 and 976, lying close to this pair on the sky.

\section{Data Acquisition and Processing}\label{processing}

\subsection{Observations}\label{acquisition}

The observations published here consist of data obtained at the VLA in
its C configuration between July and October 2005. The \hi\ spectral
line was observed with the correlator in 4IF mode using on-line
Hanning smoothing.

The observational strategy was designed to achieve the best velocity
resolution given the bandwidth needed for each galaxy. For
observations where the primary target was an edge-on galaxy (NGC 4307
and NGC 4356), we chose to overlap partially the two IFs, each one
with a bandwidth of 1.526 MHz and a spectral resolution of 24.4~kHz
($\sim 5.2$\kms), whereas for fields with main targets oriented
face-on (NGC 4411B and NGC 4492), the two IFs were centered on the
heliocentric velocity of the target, the first with a bandwidth of
1.562 MHz and a spectral resolution of 24.4~kHz, and the second using
a wider bandwidth of 3.125 MHz, but a lower frequency resolution of
97~kHz ($\sim 20.8$\kms). The goal was to search for 21 cm line
emission also from possible gaseous tidal tails, extraplanar gas or
dwarf companions in the neighborhood of the target objects. The
pointing of the field containing NGC 4492 was offset by $3\arcmin$
towards M49, due to its strong, extended radio continuum emission, in
order to avoid systematic effects due to the VLA beam squint as well
as to pointing uncertainties in individual VLA antennas
\citep{Bhat08,UC08}.

The July and August observations started in the afternoon. An
incidence in the electric system of a substation on August 12th led to
the partial loss of the observing time initially allocated for NGC
4492, which was compensated by 3 hours of diurnal observation on
September 18th. The NGC 4411B field was also observed during the
daytime on the 1st and 2nd of October (5 and 4 hrs,
respectively). Solar interference was therefore significant only for
the September and October observations. 

Each galaxy was observed for about 6 hours with an overhead of $\sim
2$ hours for calibration which included two 10-minute scans on each
day on the primary calibrator 3C286$\ =\ $1331$+$305 (J2000). The rest
of the observing sequences consisted of 30-minute scans of the target
fields interspersed with 10-minute observations of the corresponding
secondary calibrator. For the data acquired between July and
September, 3C273$\ =\ $1229$+$020 (J2000), with a flux of $\sim 32$
Jy, was used as a secondary phase and bandpass calibrator. This
calibrator was too close to the Sun during our October observations of
the NGC 4411B field. We therefore modified our strategy and observed
J1254$+$116 as a secondary calibrator. A summary of the observing
parameters is provided in Table~\ref{toribio_tab2}.

\subsection{Data Calibration}\label{calibration}

The raw UV data were reduced using the Astronomical Image Processing
System (AIPS) software package distributed by the National Radio
Astronomy Observatory.\footnote{Unless otherwise stated, all the
quoted routines in capital letters belong to this package.}

The observations of NGC 4307, NGC 4356, and NGC 4492 were calibrated
in a similar way. First, we discarded corrupted data by inspecting a
pseudo-continuum database obtained from the vector average of
visibilities for channels 4 to 60 at each time stamp ---the remaining
channels in the low and high velocity ends of the bandpass were
discarded from the beginning. The primary flux calibrator 3C286 was
used then to determine an initial bandpass as well as zeroth-order
amplitude and phase calibration. Next, pseudo-continuum images of
3C273 were obtained and subsequently used to calculate a secondary
phase calibration. The same source was the basis to determine a
secondary bandpass calibration.  This calibration was applied to the
data on the observing fields by linear interpolation of bandpass and
global phases. Finally, phase self-calibration of the data on the
galaxies was applied, and for NGC 4307 and NGC 4356, the two IFs were
joined by means of the task UJOIN \citep[channel-by-channel visibility
averages for the overlapping channels; see][]{MU08}. Regarding the NGC
4492 field, we note that during the make-up observation scheduled on
September 18th we acquired 3 hours of data that suffered from solar
interference. This forced us to reject baselines shorter than 1
k$\lambda$ for that day.

The reduction process just described was unsuitable to the
observations of the NGC 4411B field. After discarding corrupted data,
we calculated amplitude and phase calibrations from the observations
of 3C286 and 1254$+$116. A solution for the shape of the bandpass
obtained using 3C286 was applied to the whole data at the same time
that the amplitude and phase solution, which was interpolated for NGC
4411B dataset by means of simple linear connection between phases. As
for the NGC 4492 September data, we discarded spacings less than 1
k$\lambda$ to avoid the strong contamination by solar RFI.

\subsection{Continuum Emission Subtraction}\label{subsec:subtraction}

Continuum emission in our fields was estimated by imaging the vector
average of visibilities for line-free channels and subsequently
subtracted from the datacube with UVSUB.

We note that comparison of our flux measurements of the brightest
sources in each field with the corresponding flux values listed in the
NRAO VLA Sky Survey \citep[NVSS,][]{Con98} shows good agreement,
although, on average, our flux measurements are $\sim 5\pm 2\%$ larger
than the NVSS values. Given the low signal-to-noise ratio (S/N) of the
spectral signal of our targets, this does not affect our total \hi\
flux estimates. These discrepancies, however, were accounted for in
the calculation of the corresponding uncertainties.

As mentioned on \S~\ref{calibration}, for targets close to the Sun,
the images were obtained after excluding the baselines most affected
by solar contamination. However, we could not eliminate solar RFI
completely, which prompted us to make a second continuum
subtraction. The routine UVLIN was used to fit and subtract a first
order polynomial to the real and imaginary components of each
visibility through the line-free channels. Following the
recommendation by \citet*{CUH92}, we applied UVLIN also to the data
not affected by solar contamination in order to subtract any residual
continuum emission. Subsequently, we obtained the channel images and
examined the statistics of the image cubes to check for artifacts and,
in particular, to verify that the distribution of noise in our data
cubes was Gaussian-like. Finally, we proceeded to concatenate the
datasets for those objects with observations split in two different
dates (NGC 4492 and NGC 4411B).

\subsection{\hi\ Synthesis Results}\label{results}
\subsubsection{Channel Maps}\label{channels}

Image cubes were constructed for the NGC 4307, NGC 4356, and NGC 4411B
fields using robustness parameters $\Re=-1$ (closer to uniform
weighting) and $\Re=0.7$ (closer to natural weighting). For all
galaxies but NGC 4492, we present the results for $\Re=0.7$ as it
provides the best compromise between the S/N and resolution when data
have full UV-coverage \citep{Bri95}.

The low S/N of the data on NGC 4492 forced us to smooth and taper the
observations to improve our sensitivity. Channel maps were obtained
from IF 2 (spectral resolution $\sim 20.8$\kms) by using a robustness
parameter $\Re=1$, which results in lower noise levels and a wider,
but still acceptable, synthesized beam size. A Gaussian taper (with a
9 k$\lambda$-width at the $30\%$ level both in the U and V directions)
additional weighting was applied on the visibilities to lower the
contribution of the long-baseline datapoints.

In all cases, the channel images were CLEANed and the clean components
restored with a Gaussian beam similar to the synthesized beam. The
characteristics of the deconvolved images are summarized in
Table~\ref{toribio_tab3}.

\subsubsection{Moments}\label{moments}

The cleaned image cubes were used to calculate total \hi\ flux images,
as well as first and second velocity moments of the 21 cm
emission. 

We decided whether to keep or not a pixel in the integration by
examining a spatially and frequency smoothed version of the
datacubes. The spatial smoothing was done by convolving with a
Gaussian kernel, whereas a Hanning smoothing was applied in
velocity. We selected those pixels from the original datacubes that
were above the $3\sigma$ level in the smoothed counterpart, except for
NGC 4492, for which the $2\sigma$ level was used (otherwise, almost no
signal was left from the galaxy). All flux with absolute value above
$0.5\sigma$ was integrated along the velocity axis to obtain the total
\hi\ map, and intensity-weighted first and second order moments were
computed. Finally, the \hi\ intensity map was corrected for primary
beam attenuation and scaled to the column density $\nhi$ of the gas,
assuming optically thin \hi. When estimating the errors in the latter
map, we scaled the rms noise in the channel maps to the square-root of
the mean number of adjacent channels that contributed to the total
intensity image, and then added a $5\%$ independent uncertainty
arising from calibration and correction for primary beam attenuation.

The diameter of the \hi\ disks was measured at a column density of
$10^{20}$ atoms~cm$^{-2}$ (no correction for beam smearing was applied).
The reported uncertainties take into account variations in position
angle of major axis as well as the correlation introduced by the 
synthesized beam (Table~\ref{toribio_tab4}).

\subsubsection{Global \hi\ Profiles}\label{globalprofiles}

In order to determine if we have recovered all of the flux from
single-dish measurements, simulated line profiles were derived. The
latter were calculated by integrating over the spatial axes for each
channel the primary-beam attenuation-corrected emission used in the
moment calculation (\S~\ref{moments}).

We have measured $\wtwenty$ and $\wfifty$, the profile width at the
$20\%$ and $50\%$ of the peak intensity, respectively. For those cases
in which a clear double-peaked profile is found (VCC 740, NGC 4411A and NGC
4411B), the peak fluxes on both sides were considered separately when
calculating the linewidths. In the remaining cases we used the overall
peak flux. The \hi\ linewidths were subsequently corrected for
broadening effects due to the finite spectral resolution of the
instrument by following the considerations of \citet{VS01}. The
adopted broadening corrections for the cubes with a spectral
resolution of $5.2$\kms\ were $\delta\wtwenty = 0.86$\kms\ and
$\delta\wfifty =0.56$\kms, whereas when the resolution was 20.8\kms,
we adopted $\delta\wtwenty =12.0$\kms\ and $\delta\wfifty
=7.88$\kms. 

The heliocentric systemic velocity, $v_\mathrm{sys}$, was 
derived as the average of velocities of channels at 20 
and 50\% of the peak flux of the profile, and the total
\hi\ flux was obtained by integrating the 21 cm line 
profiles along the velocity axis.

Since some of the profiles have low S/N, we decided to 
measure the kinematic parameters defined above by running Monte
Carlo simulations for each one of the profiles. The values 
quoted in Table~\ref{toribio_tab4} correspond to the mean and 
1-$\sigma$ deviation of the distribution of measurements from
one thousand random realizations of the profiles by 
taking into account the flux in each channel and its rms
error. The latter was estimated from the rms noise in 
the channel maps and the correlation introduced by 
the synthesized beam. The quoted error in the 
total \hi\ flux has been estimated by adding in 
quadrature the error estimates from the Monte Carlo
technique and a $5\%$ uncertainty arising from the
calibration and the correction for primary beam attenuation.

We find, in general, good agreement with single-dish observations
(Table~\ref{toribio_tab4}). Especially remarkable is the close match
between the shapes, linewidths, and total flux densities of our
VLA \hi\ line profiles and their counterparts in the ongoing
Arecibo Legacy Fast ALFA (ALFALFA) extragalactic \hi\ survey
\citep{Gio05}, which recently released the results for the strip of
the Virgo Cluster region where our targets are located \citep{Ken08}.

\subsubsection{Position-Velocity Diagrams and Rotation Curves}\label{rotcurves}

Position-velocity (PV) diagrams along the major axis of each target
were also obtained by taking slices of the data cubes through the
optical centers and estimating their position angle from the moment
maps or from the rotation curves in those cases where this was
possible.

Major-axis velocities as a function of angular radius were exclusively
derived for the two components of the pair NGC 4411A/B and the dwarf
spiral VCC 740, since these were the only targets detected with a high
S/N (see next section). To infer the rotation curves we have fitted
both a Brandt model and the standard iterative tilted-ring algorithm
\citep{Beg89}.

\section{Case by Case Analysis of the \hi\ Synthesis
  Data}\label{analysis}

Table~\ref{toribio_tab4} summarizes the results inferred from our
analysis of the VLA observations for all the galaxies detected. In the
printed version of the manuscript we include a complete graphical
layout of the NGC 4307 field
(Figs.~\ref{channels_n4307}--\ref{synthesis_n4307}). Images for the
rest of our VLA pointings are available in the electronic version of
the manuscript.

We now comment on the properties of the target galaxies from our VLA
observations.

\subsection{NGC 4307}

The \hi\ gas disk of NGC 4307 has very small dimensions and is found
only deep within the optical disk, in agreement with the strong
gas-deficiency estimated for this galaxy. The deficiency parameter
$\df$ (eq.~[\ref{def}]) measured from our VLA flux (see
Table~\ref{toribio_tab4}) shows a pretty good consistency with the
value of 1.41 one can infer from the observed \hi\ flux provided by
\citep{Ken08}. We remark that this new value is also in agreement with
the extremely \hi-deficient status reported in previous works (e.g.\
\citeauthor{Sol02}; \citealt{Gav05}).

The $0$th-order moment map shows that the maximum of the 21 cm
emission is displaced SE with respect to the optical center, in a
direction nearly perpendicular to the major axis, with an excess of
emission at the approaching side. This asymmetry shows up also in the
global \hi\ profile, which has a peaked appearance indicative of
centrally concentrated gas. The estimated offset of $\sim 7\arcsec$ is
less than half of both the synthesized beam size and the mismatch
expected to arise from edge-on ram-pressure stripping in hydrodynamic
simulations \citep{Kro08}.

There are no other pieces of evidence susceptible to being interpreted
as environmental effects down to the sensitivity limit of the
measurements. Moreover, the galaxy is too inclined to allow one to
disentangle whether the observed asymmetries can be ascribed to
noncircular motions associated with a spiral arm.

The PV diagram along the major axis is still rising at the edge of the
measured \hi\ distribution, as if the gas was rotating almost as a
pure solid body. Our VLA data is compatible to a large extent with the
H$\alpha$ rotation curve derived by \citet*{Gav99}, which shows a hint
of a turn over on both sides of the galaxy. We discuss in
Section~\ref{tf} the risks of using extremely \hi-deficient galaxies
like this one in a TF analysis.

\subsection{NGC 4356 and VCC 740}

The properties of the neutral hydrogen in NGC 4356 resemble those of
NGC 4307. This is a nearly edge-on galaxy that like the former one,
and even more strongly so, has a very small gas distribution compared
with its optical dimensions. The PV diagram along the major axis also
rises steeply and shows no signs of turning over, as in the H$\alpha$
rotation curve measured by \citet*{Gav99}. There is similarly a
misalignment between the distribution and motion of the atomic gas
and the optical disk. As in the former case, this offset is less than
half the synthesized beam size and, hence, relatively small and not
necessarily indicative of ram-pressure pushing.

For this galaxy, our VLA linewidth and flux values are somewhat
smaller that the most recent estimates by \citet{Ken08}. Given the low
S/N of this galaxy, the observed difference in total fluxes, which is
only of tenths of a mJy\kms, does not necessarily imply that some flux
has been lost due to the missing short baselines in our synthesis
aperture images. Instead, the discrepancies can be the result of 
low-baseline ripples that seem to affect the single-dish profile.

VCC 740, another highly inclined galaxy detected in this same
pointing, has, in contrast, quite a strong S/N and a total \hi\ flux
that is about the same that in NGC 4307. Its \hi\ map suggests that
the neutral gas has a sharp cutoff at the optical radius of the disk
on the approaching side, while the other side has a more gradual
falloff and is somewhat more extended. In spite of giving the
impression that part of the gas might be missing, both ours and the
ALFALFA flux measurements actually result in a negative deficiency
parameter $\df$ of $\sim -0.2$, indicative of a perfectly normal \hi\
content.

The gas velocity field of VCC 740, on the other hand, exhibits some
warping, as well as inner contours parallel to the minor axis,
indicating that $V_{\mathrm{rot}}(R)\propto R$, as the PV diagram
shows. Further out on the SE side, the contours show the classical
$V$-shape, suggesting that at the sensitivity limit of the
measurements the rotation speed is just about to become nearly
constant. In the rotation curve model fits, the warp of the velocity
field increases the position angle on the external part of the
disk. Both Brandt and tilted-ring models yield an estimated
inclination of $\sim 70 \degr$ for the internal disk region that drops
to $\sim 45 \degr$ on the outside. The fact that the velocity field of
this galaxy shows a clear rotation pattern reinforces its
morphological classification as a dwarf barred disk instead of the IB
type assigned in the LEDA. The angular resolution of our VLA
measurements, however, is insufficient to observe the effects of the
bar on the gas velocities. No signs of interaction are found between
VCC 740 and NGC 4356.

\subsection{NGC 4411A/B, VCC 933 and VCC 976}\label{NGC 4411A/B}

Contrarily to what has happened with NGC 4307 and NGC 4356, the two
low surface brightness spirals NGC 4411A and B have produced
integrated \hi\ fluxes that are inconsistent with the values quoted in
\citeauthor{Sol02}, which were estimated in earliest mapping attempts
done with the Arecibo antenna \citep{Hay81}. Discrepancies with this
and other old single-dish measurements \citep[e.g.][]{Hof89} are
ascribed, however, solely to the amplitudes, as the shapes of the
global \hi\ profiles look very similar.

Our new data, reinforced by the newest single-dish measurements done
at Arecibo by the ALFALFA team \citep{Ken08}, imply that NGC 4411A and
NGC 4411B loose their initial status of objects with moderate and
strong \hi\ deficiency, respectively (see Table~\ref{toribio_tab1}),
to be both reclassified as galaxies with quite a normal gas abundance, 
in accordance with the visual impression obtained from the \hi\
contours overlaid on the Digitized Sky Survey (DSS) image.

These maps show that these two galaxies have \hi\ distributions
extending beyond their optical disks, which for NGC 4411A reach up to
nearly twice the optical radius, except on the NE side where the \hi\
contours appear compressed. In this latter galaxy the \hi\ is
concentrated in a ring with two important regions of emission that
emanate perpendicularly from the ends of the bar. The $0$th-order map
of NGC 4411B shows an even wider major ring-like structure with a
noticeable excess of emission at its northern half near the outer
edge. We do not find important displacements from the optical disks,
the most remarkable feature being the presence of a depression in the
center of both galaxies, as found in other LSB galaxies
\citep*{dBMvH96}.

We fail in detecting neither gas bridges nor significant intergalactic
\hi\ signals between these two targets. Given that both galaxies show
normal disk emission with only mild alterations of the symmetry,
comparable to those seen in more isolated objects \citep{Kor00}, we
feel compelled to classify the system NGC 4411A/B as a visual pair
(see also \S~\ref{tf}).

Regarding the kinematics of the neutral gas, we note that the PV
diagram of NGC 4411B shows a steep rise well within the stellar disk
(of small amplitude given its near face-on orientation) followed by a
sharp bend towards a flat part, ---with indications of a modest
decline on the receding side at the largest radii where the \hi\ is
detected---, consistent with the shape of the radial velocity contours
and with the well behaved double-horned global profile. In NGC 4411A,
the turnover in the rotation velocity is not complete and the 21 cm
line profile is affected by a larger asymmetry. The observed behavior
of the PV diagrams is typical of objects with a compact distribution
of their luminous matter. 

For these two galaxies we have gone a step further and inferred also
major-axis rotation velocities with the aim of estimating the
orientations of the gaseous disks. The apparent inclination inferred
from the tilted-ring technique for NGC 4411A is $29^{+5.2}_{-3.7}$
degrees, while for NGC 4411B we get $i=26^{+4.4}_{-4.7}$
degrees. Brandt curve fits to the whole velocity fields result in
similar inclinations of $\sim 27\degr$ and $\sim 28\degr$,
respectively. Yet it shouldn't be forgotten that the bumps in the \hi\
line data associated with spiral arms and, especially, the near
face-on orientation of these disks, make rotation curve model fits to
the velocity fields uncertain and, caveat lector, unable of giving a
precise inclination angle. In the next section, we discuss the
problems arising from the derivation of this parameter in
low-inclination galaxies and provide alternative estimates based on
optical images.

Our VLA observations have not detected 21 cm line emission coming from
any of the two dwarf elliptical galaxies VCC 933 and VCC 976 also
present in this field.

\subsection{NGC 4492}

In spite of the strong image-degrading effects of the Sun for the
second half of the observations of NGC 4492, we succeed in detecting
its \hi\ signal at a quality level comparable with the previous
single-dish observations carried out by \citet{HG86} and
\citet{Hof89} ---and not too different from the one achieved by the
ALFALFA measurements, which have a S/N of 4.6.

In this galaxy, the detected neutral hydrogen is located within the
optical radius and shows an important elongation in the SE-NW
direction, almost perpendicularly to the galaxy major axis. The \hi\
distribution is strongly asymmetric, with the peak of the \hi\
emission shifted some $30\arcsec$ to the East from the optical center
of the galaxy. This gives rise to a synthesized line profile with
decreasing flux toward the approaching side. By integrating the
latter, we find a total \hi\ flux about a $30\%$ smaller than the
ALFALFA's value (see Table~\ref{toribio_tab4}).  Again, one may wonder
whether possible flux losses in our estimation arising from too a
strict rejection of the short baselines could explain this
difference. In this respect, we note that during the application of
the Gaussian UV-taper weighting to obtain the final datacubes some
testing done varying the width of the tapering function showed that
the resulting total \hi\ flux was not significantly affected.
Therefore, given the low S/R of the detections, the discrepancies in
total flux and linewidths with respect to singledish measurements
could have been originated by low-level baseline ripples that seem to
affect the ALFALFA profile.  With this caveat in mind, our VLA
measurements assign a new \hi\ deficiency parameter of 1.21 to NGC
4492, which indicates that the gas content for this galaxy could be
less than $6\%$ of the expectation value.

The asymmetries on the spatial distribution of the gas for this galaxy
are reproduced in the \hi\ dynamics. Thus, the velocity map of NGC
4492 reveals a possible displacement of the dynamical center from the
light center (consistent with that observed in the $0$th-order map),
while on the PV diagram along the major axis most of the emission is
found coming from the receding side. The low S/N of the datacubes, the
small size of the \hi\ disk with respect to the beam size, as well as
a not too favorable orientation of this galaxy, prevent any attempt to
fit a rotation curve model to the velocity data.

\section{TF-based Distance Estimates to our Galaxies}\label{tf}

The most striking aspect of the Virgo's galaxies selected in the
present study is not their high deficiency of neutral gas, but the
possibility that this has been attained outside the cluster
environment. While in the outskirts of clusters gas stripping can
happen by galaxy-galaxy interactions in infalling groups or through
collisions with lumps of intergalactic gas \citetext{see, for
  instance, \citealt*{CK06,CK08}}, strong \hi\ deficiencies in the
periphery of clusters are expected to be an exception to the rule.

The classification by \citet{San04} of a spiral galaxy as an \hi\
outlier relied on its TF distance. The latter was inferred from the
disk's maximum rotation speed $V_{\mathrm{rot}}$, ---which is expected
to be a proxy to the total mass of the galaxy and, therefore, also to
its intrinsic luminosity---, measured via the width of the \hi\
spectral line. It is then of fundamental importance to assess the
feasibility of this technique in galaxies like ours which have gaseous
disks deeply altered.

From \citet{Guh88} to \citet*{CKH08} the literature is full of TF
studies highlighting the risks of using the gas kinematics in the
determination of distances to \hi-deficient galaxies. The most straight
reasoning being that truncated gas disk measurements could underestimate
the rotation velocity of a galaxy and, therefore, its mass, biasing
low the derived radial distance. However, depending on the interaction
mechanism and its geometry, the \hi\ that does not get dislocated may
also lost temporarily its equilibrium within the global galactic
potential ---externally induced disturbances on the kinematics of
disks are erased in about 1 Gyr \citep[e.g.][]{Dal01}, a time during
which two interacting galaxies can move hundreds of kiloparsecs
apart. These effects, as well as induced noncircular or nonplanar gas
motions, which may even lead to an overestimation of the true
$V_{\mathrm{rot}}$ \citep{Kro07}, possible changes in the observed
luminosity resulting from alterations in the star formation rate or,
simply, the fact that even for undisturbed galaxies in many cases
there is evidence of noncircular motions in their central regions
\citetext{for instance, due to bars; \citealt*{Val07}}, could make the
application of the TF technique in strong \hi-deficient galaxies
totally inefficient.

All this drives us to regard critically the distance estimates of the
three galaxies in our sample that exhibit severely truncated gas
disks: NGC 4307, NGC 4356, and NGC 4492. Certainly, we have not found
evidence of a recent gravitational encounter in our 21 cm line imaging
data in the form of gaseous tails and bridges for any of them. Nor the
closeness of the systemic velocity of the first two galaxies to the
mean cluster velocity supports a recent ram-pressure event inside the
core. Yet the fact we do not detect a flat part in 
their rotation curves, as well as the irregularities in the spatial 
distribution and kinematics of the \hi\ evidenced in the VLA maps 
of these objects do not allow us to state confidently that 
the dynamical equilibrium of 
the neutral gas has been fully restored after the removal
event. Neither can we assert that their luminosities have not been
affected. Therefore, we believe that it is not legitimate to use the
TF technique to derive the radial distance to any of these three
galaxies and that, consequently, their published TF distance
measurements might well be largely in error.

Compared to the previous objects, NGC 4411B and its close neighbor NGC
4411A exhibit regular and symmetric gas velocity fields, with flat
extended outer parts that appear to satisfy the basic tenet that
underlies a TF study. In this case, however, the attempts of
estimating the radial distance to these two galaxies are thwarted for
an unfavorable viewing angle. At low apparent inclinations ($i\lesssim
40$--$45\degr$), estimates of the orientation of disks become more
uncertain, led to deprojected quantities with divergent errors as a
face-on orientation is approached, and are skewed towards larger
values by nonaxysimmetric features in the images \citetext{see
  \citealt*{AB03} and references therein}. As a result, the total
fractional error in the radial distance for low-inclination galaxies
largely exceeds $15\%$, a value usually adopted as representative of
the typical uncertainty in the distance arising from good-quality TF
data.

This situation is illustrated in Figure~\ref{tf_distances}, where we
show the radial distances to NGC 4411A and B reported in some of the
TF catalogs used by \citeauthor{Sol02}. Our synthesized \hi\ line
profiles have been used to estimate the intrinsic rotational
velocities, which we have calculated by exactly following the same
prescriptions adopted in the referenced studies that in some cases
apply non-null turbulent motion corrections. The uncertainty resulting
from inclination measures is shown by a vertical bar whose extent
is set by the most extreme values of this angle ever assigned to our
galaxies in the literature (which range from $\sim 20\degr$ to $\sim
55\degr$). Open diamonds in the plots indicate radial distances
published in the cited references. It is obvious from this figure that
the inability to infer accurate inclinations prevents us from
establishing the location of both galaxies along the LOS \emph{within
  the entire Virgo Cluster region}.

We have also depicted in Figure~\ref{tf_distances} the \hi-based
radial distances inferred from our own measurements of the inclination
of these galaxies (Sec.~\ref{NGC 4411A/B}). Double pointed arrows
encompass the ranges of distances corresponding to the values of
inclinations and errors derived from inspection of the residuals of
our tilted-ring model fit to the \hi\ velocity fields. The well-known
limitations of the weighting scheme included in the modeling of the
\hi\ rotation curves when it comes to measuring inclinations below
$40\degr$ \citep{AB03}, have led us to make as well independent fits
to the orientation parameters of these disks from Sloan Digital Sky
Survey (SDSS) images. We have used the package GALFIT \citep{Pen02} to
decompose the r-band images of both galaxies into several
components. The sky contribution apart, we have fitted a bulge plus an
exponential disk to the image of NGC 4411B, while for NGC 4411A a bar
component has been added too. We have followed \citet*{BdV81} and
adopted intrinsic axis ratios of 0.18 for NGC 4411A (SBc) and of 0.13
for NGC 4411B (Scd). All this gives estimated inclinations of about
$24\degr$ for the stellar disk of NGC 4411A and of $\sim 16\degr$ for
the one of NGC 4411B. The uncertainties in these values are difficult
to determine, as the analysis of the optical images involves a large
number of parameters. Resulting distances are indicated in the plots
by asterisks. Comparison with the results from the \hi\ shows that for
nearly face-on disks in the Virgo Cluster region inclinations
differing by $\sim 10\degr$ can lead, depending on the TF relationship
adopted, to differences in distance exceeding 10 Mpc. The very
uncertain radial distances and the lack of unmistakable signs of an
ongoing interaction do not make it possible to assert that these
galaxies are physically connected in spite of their proximity in
$z$-space.

Another feature of Figure~\ref{tf_distances} that draws one's
attention is the existence of considerable author to author
fluctuations in the estimated TF radial distances that cannot be just
ascribed to the uncertainties in the observational parameters entering
this relationship. For highly inclined galaxies, such discrepancies
can be traced back mostly to the adopted TF template, which for a
given passband can show systematic variations among different authors,
even when similar samples of calibrators are used. We note, for
instance, that small datasets are rather sensitive to the trimming of
the data and, hence, to the, somewhat subjective, identification of
the most deviant measurements. Besides, differences on the fitting
methods, or on the adopted data weighting, as well as morphological
and incompleteness biases \citep[see, for instance,][]{Gio97}, may
give raise to significant variations in the slope and zero-point
coefficients of the TF relation. As shown in
Figure~\ref{tf_templates}, where we compare the $B$-band TF template
relations defined in the studies of the Virgo Cluster considered by
\citeauthor{Sol02}, the absolute magnitude assigned to a given galaxy,
regardless of its inclination, can vary up to $\sim 1.5$ mag depending
on the calibration adopted. This exceeds by far the typical error of
$\sim 0.1$--0.2 mag usually assigned to the zero-point calibration of
individual TF templates.

\section{Summary and Conclusions}\label{conclusions}

This is the first 21 cm synthesis survey of spiral galaxies ever made
in which the targets have been specifically chosen on the basis of
their expected dearth of cold gas. It has been motivated by the
detection, in previous investigations of the neutral gas content on
spirals in the Virgo Cluster region, of a significant number of
severely \hi-deficient disks supposedly located, according to their TF
distance estimates, beyond the maximum rebound radius galaxies can
bounce after infall. According to this location, these galaxies could
not owe their \hi\ deficiency to interactions within the cluster
environment.

Our high-sensitivity VLA observations have been aimed at
characterizing in detail the spatial distribution and kinematics of
the neutral gas in four galaxies suspected of being \hi-outliers, in a
first effort to gain a better understanding of the origin of this
class of objects. At the same time, these synthesis observations in
the 21 cm line have provided direct evidence of the risks involved in
the application of the TF relationship to disturbed or nearly face-on
disks, which can render the derived distances unreliable.

We have detected a total of six galaxies within the four fields
initially selected. The main conclusions of the analysis of our VLA
data are as follows:

1) We confirm the strong \hi\ deficiency of three of our four main
targets, NGC 4307, NGC 4356, and NGC 4492 (inferred \hi\ contents are
a factor $\gtrsim 20$ lower than their corresponding standard values),
which is pronounced through the reduced extension of the gaseous
disks, a characteristic typical of ram-pressure-stripped galaxies. In
contrast, we find that the integrated \hi\ fluxes of our fourth
target, NGC 4411B, and its companion, NGC 4411A, are $\sim 2$--$3$
times larger that the old single-dish values used to estimate their
\hi\ deficiency in \citeauthor{Sol02}. Our measured VLA fluxes,
---which for all our targets are compatible with those inferred from
the new, sensitive ALFALFA extragalactic \hi\ survey---, indicate that
the \hi\ contents of these last two galaxies deviate less than
$1\sigma$ from normalcy. This is consistent with our observation that
their \hi\ disks extend beyond their optical counterparts, so that
only the outermost portions of the cold gas distributions have been
affected, if at all. A sixth galaxy with a healthy amount of cold gas,
the dwarf spiral VCC 740, has been detected in the field of NGC 4356.

2) Visual inspection of the images of the most gas-deficient galaxies
has revealed signs of asymmetries and lopsidedness, as well as small
offsets of the dynamical centers with respect to the optical
ones. These are all suggestive, albeit not conclusive, indications of
possible gravitational interactions and/or ram-pressure effects, as
deviations from flat, axisymmetric disks are also known to prosper in
isolated galaxies. This, and the fact that we have not found evidence
of gaseous tails or bridges within the limit we have been able to
trace the \hi\ ($\sim 3$--$5\times 10^{19}\ \mathrm{cm}^{-2}\
\mathrm{channel}^{-1}$ at the $3\sigma$ level) appear to indicate that
none of the galaxies investigated has undergone \emph{recent}
gravitational interactions. This means, in particular, that our VLA
observations reinforce the classification of NGC 4411A/B as a virtual
pair in spite of their closeness on the observational space phase.

3) Our three targets with highly truncated gas disks exhibit rotation
velocities that are still rising at the last measured points.
Moreover, the observational evidence gathered do not allow us to
assert with complete confidence that the gas remaining tied to the
disks has regain dynamical equilibrium, nor the extent to which the
luminosity of these \hi-deficient galaxies could have been
affected. The classification of these objects as \hi-outliers could
therefore simply obey to the inefficient estimate of their radial
distances by means of the TF relationship. The fourth target, NGC
4411B, as well as it space phase neighbor NGC 4411A, show, in
contrast, extended gas disks with regular and symmetric velocity
fields. In spite of being galaxies presumably in virial equilibrium,
their TF-based distances are also problematic because of their nearly
face-on apparent orientation, which results in the inability to
determine accurate inclinations. This translates into a considerable
uncertainty ---larger than the resolution necessary to determine
unambiguously the region (infall or cluster) where a galaxy belongs---
when it comes to placing these two galaxies along the LOS within the
Virgo Cluster region.

Aperture synthesis observations in the 21 cm line like the ones
presented here are fundamental for probing the impact of cluster
residency on the spiral population. Further insight into the
identification of the physical processes disturbing the disks can be
gained by supplementing this type of data with multifrequency
observations.

\begin{acknowledgements}
  This work was partly supported by the Direcci\'on
  General de Investigaci\'on Cient\'{\i}fica y T\'ecnica, under
  contracts AYA2006--01213 and AYA2007-60366. M.C.T.\ acknowledges
  support from a fellowship of the Ministerio de Educaci\'on y Ciencia
  of Spain. We are grateful to all the people and institutions that
  have made possible the LEDA (\texttt{http://leda.univ-lyon1.fr/}),
  GOLDMine (\texttt{http://goldmine.mib.\linebreak[0]infn.it/}), DSS
  (\texttt{http://archive.stsci.edu/cgi-bin/dss\_form}), and SDSS
  (\texttt{http://cas.\linebreak[0]sdss.org}) databases.
\end{acknowledgements}


\input{toribio_figures}




  \clearpage 
  \input{toribio_tab1_selection}

  \clearpage
  \input{toribio_tab2_observations}

  \clearpage
  \input{toribio_tab3_deconvolution}

  \input{toribio_tab4_analysis}

\end{document}

%% file: toribio_figures.tex
\onecolumn

\begin{figure}
  \epsscale{0.8}\plotone{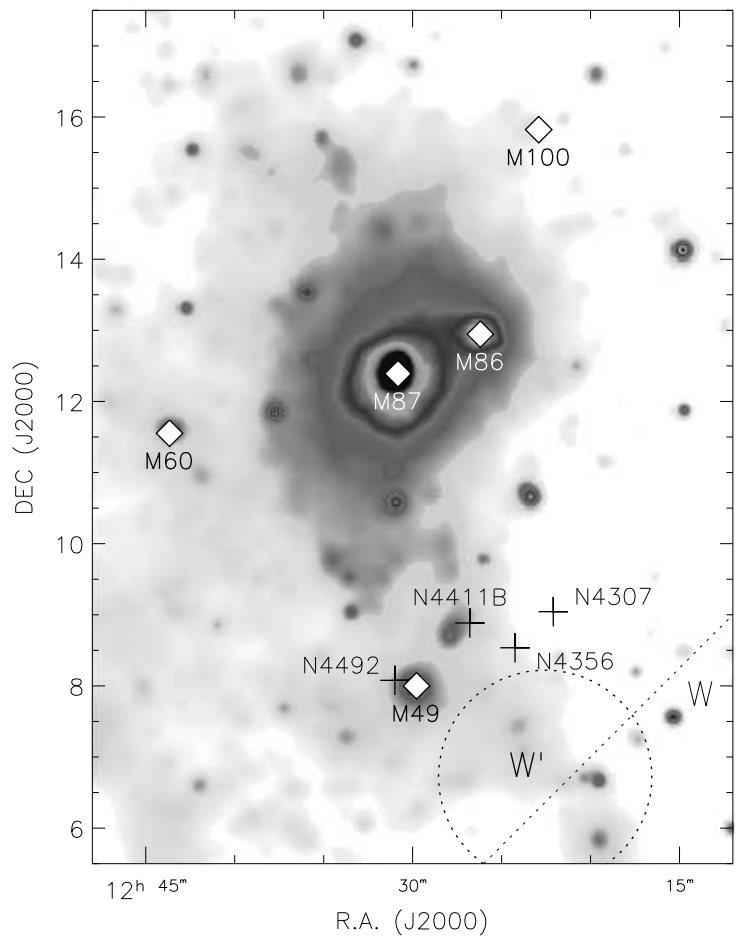} \figcaption{Sky
    distribution of our 4 main targets. The positions of five dominant
    Virgo cluster galaxies are marked by open diamonds (\emph{top to
      bottom:} M100, M86, M87, M60, and M49). Overlaid on the figure
    it is the greyscale map of the X-ray emission in the cluster from
    the \emph{ROSAT All-Sky Survey} in the 0.4--2.4 keV band \citep{Boh94}.
	The location of W and W$\prime$ clouds is also shown.\label{location}}
\end{figure}


\begin{figure}
\figurenum{2}
\plotone{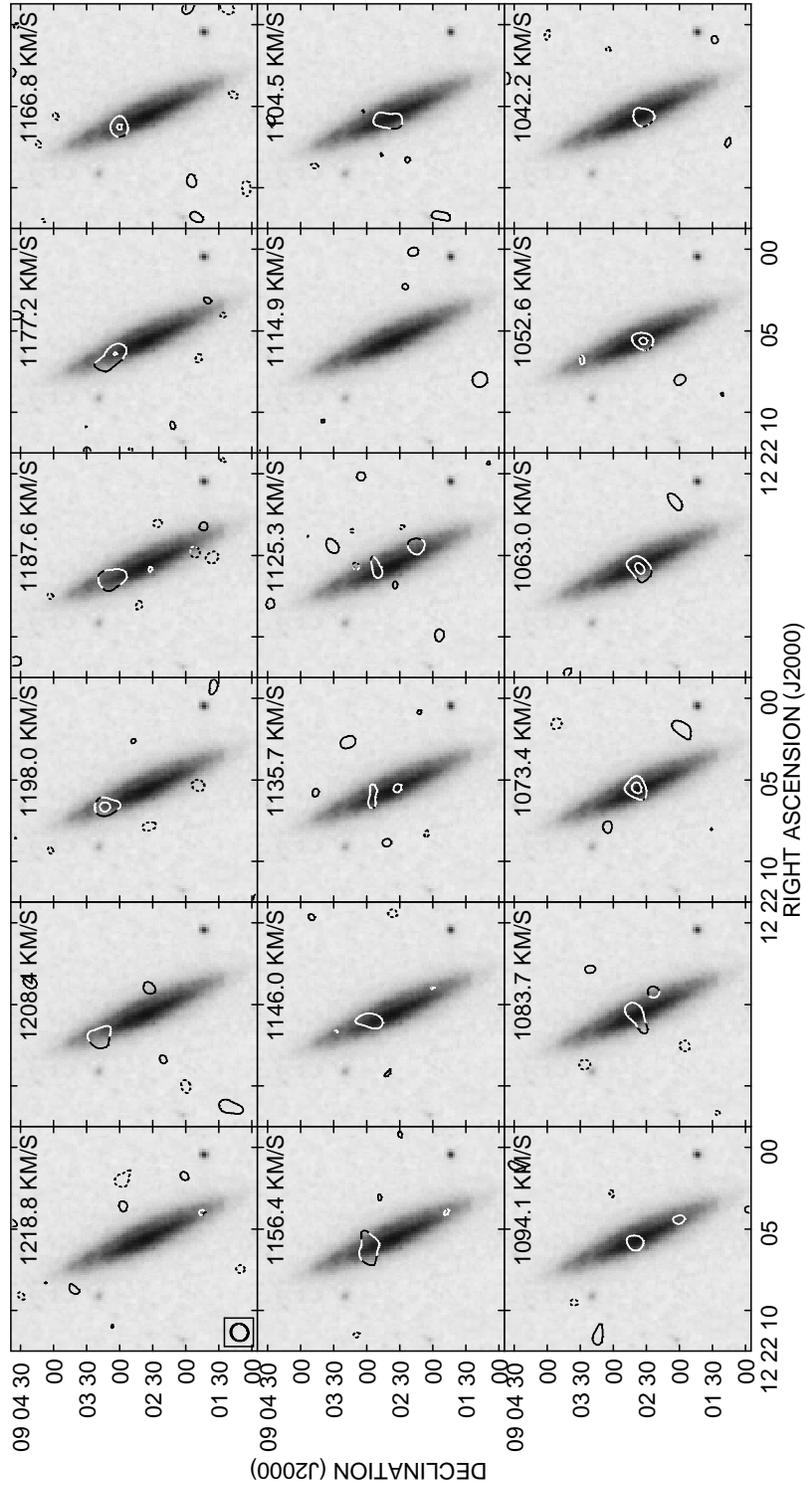}
\caption{\hi\ channel images for NGC 4307 overlaid on the DSS image of the galaxy. The maps were obtained by averaging the channels in pairs. Only channels containing line emission are plotted, bracketed by two noise channels. The heliocentric velocity of each panel is given in the upper right corner. The contours are drawn at ($-2$~[absent], $-1$~(dashed), 1, 2) $\times 3\sigma$, where $\sigma \sim 0.42$ mJy beam$^{-1}$ is the rms noise level (see Table~\ref{toribio_tab3}).}
\label{channels_n4307}
\end{figure}

\begin{figure}
\figurenum{2}
\plotone{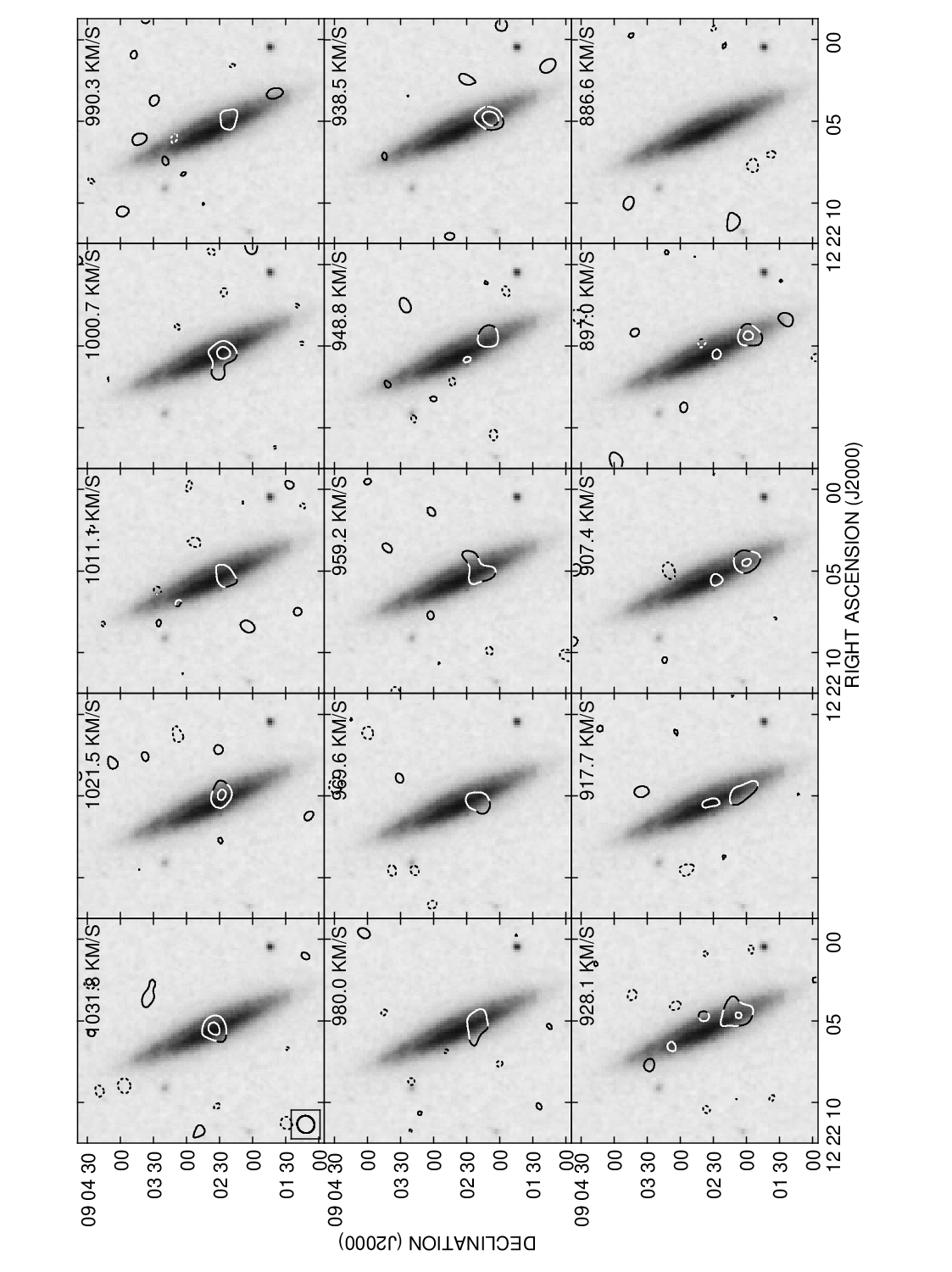}
\caption{Continued.}
\end{figure}


\begin{figure}
\figurenum{3}
  \figcaption{\hi\ synthesis results for NGC 4307. {\it (a)} \hi\
    column density contours overlaid on the optical image of the
    galaxy obtained from the DSS. The contours levels are
    ($-2$~[absent], $-1$~(dashed), 1, 2, 4, 8) $\times\ $ 31 mJy
    beam$^{-1}$\kms. The size of the synthesized beam is plotted in the
    lower left corner. {\it (b)} Linear gray-scale map of the \hi\
    column density.  The grayscale range is 31--276 mJy
    beam$^{-1}$\kms. {\it (c)} Velocity field of the \hi\ with contours
    and grayscales ranging from 899 \kms\ (light gray) to 1261 \kms\
    (dark gray) in increments of 20.8 \kms. The adopted heliocentric
    systemic velocity is indicated by a thick black contour. {\it (d)}
    \hi\ position-velocity plot along the major axis. The data have
    been averaged in velocity (joining the channels in pairs) and
    spatially along a $8\arcsec$-wide strip through the optical center
    of the galaxy with a position angle $\Gamma=25\degr$ (East of
    North). Contour levels are at ($-\sqrt{2}$~[absent],
    $-1$~(dashed), 1, $\sqrt{2}$, 2) $\times$ 1.44 mJy
    beam$^{-1}$. The gray scale is linear in the range [0.96--3.11]
    mJy beam$^{-1}$. A pair of vertical arrows are drawn to indicate
    the extent of the optical disk.{\it (e)} Integrated
    \hi\ line profile from our VLA observations (solid line) and ALFALFA (dotted line). The vertical arrow
    indicates the systemic velocity from the VLA profile. 
\label{synthesis_n4307}}
 \begin{tabular}{cc}
\includegraphics[width=0.5\textwidth]{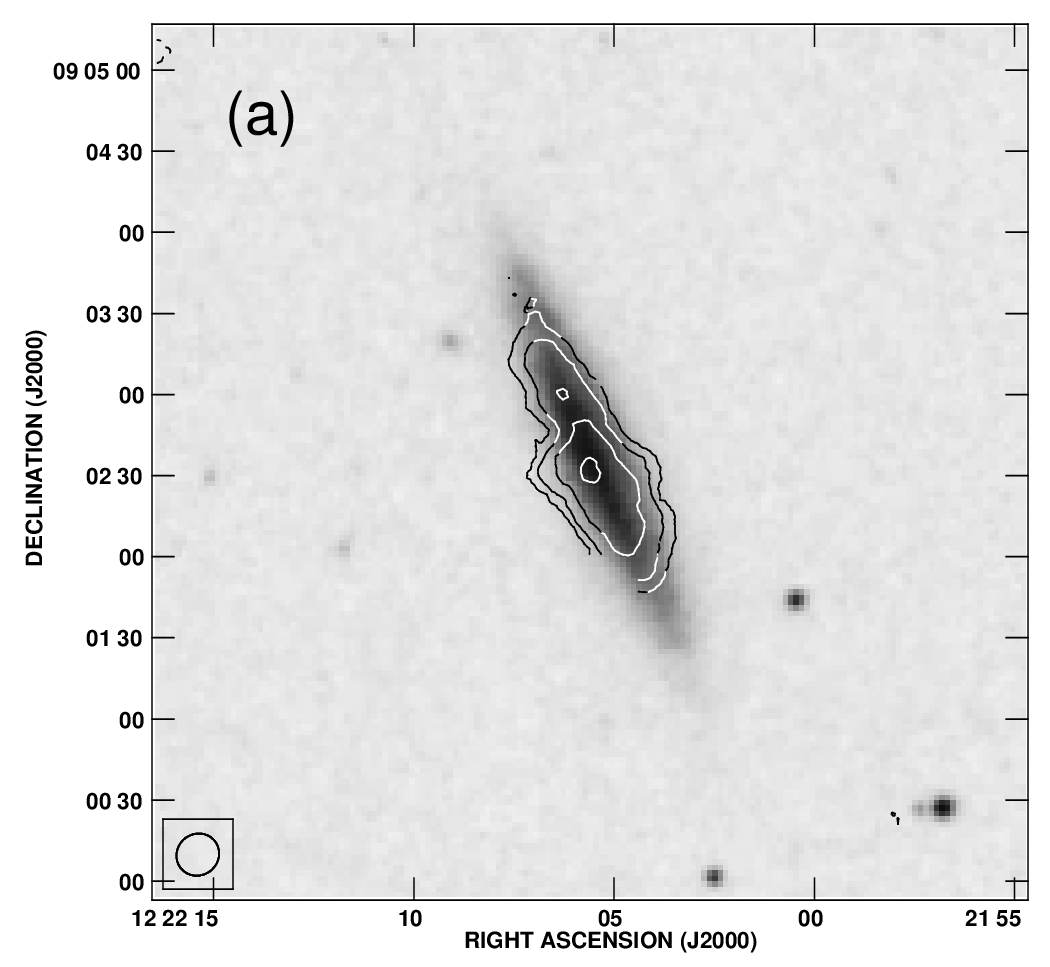} &
\includegraphics[width=0.5\textwidth]{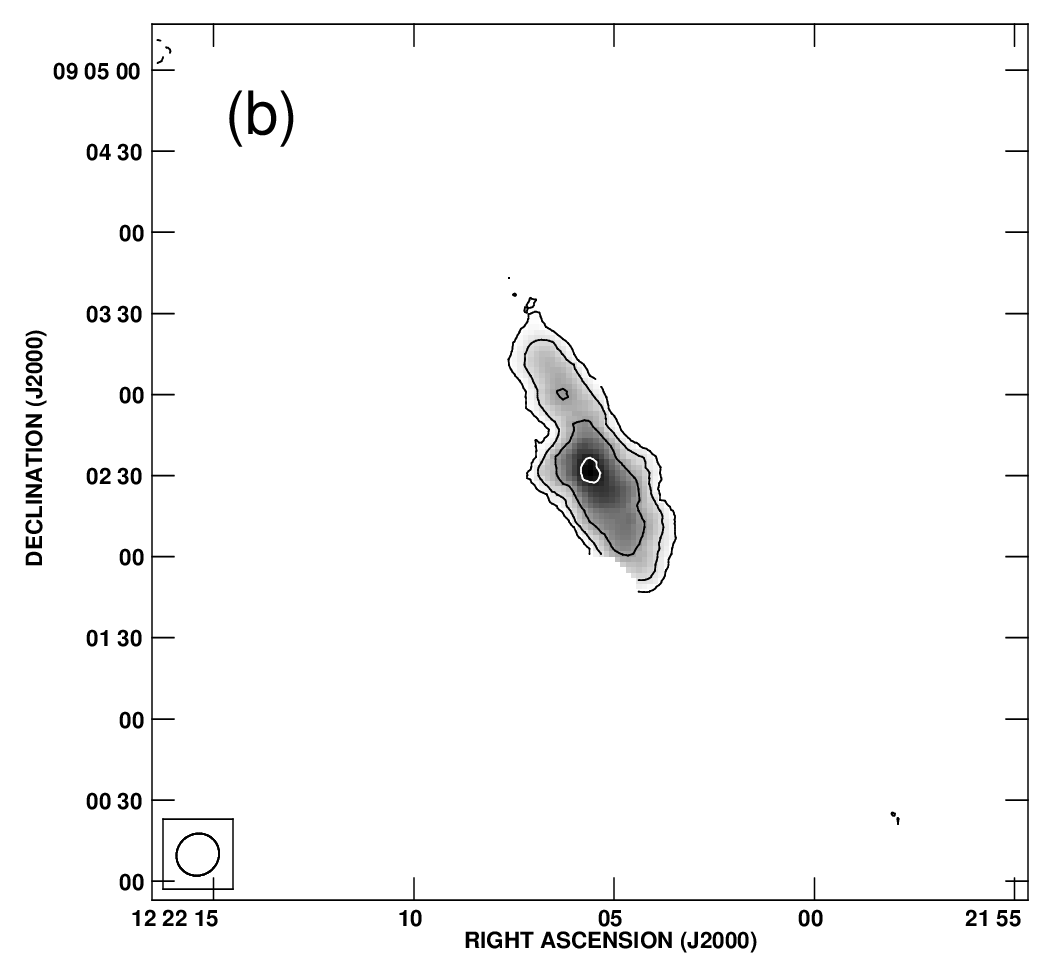}\\

 \end{tabular}
\end{figure}

\begin{figure}[t]
\figurenum{3}
  \begin{tabular}{cc}
\includegraphics[width=0.5\textwidth]{f3_1c.ps} &
\includegraphics[height=0.45\textwidth]{f3_1d.ps}\\
\includegraphics[width=0.5\textwidth]{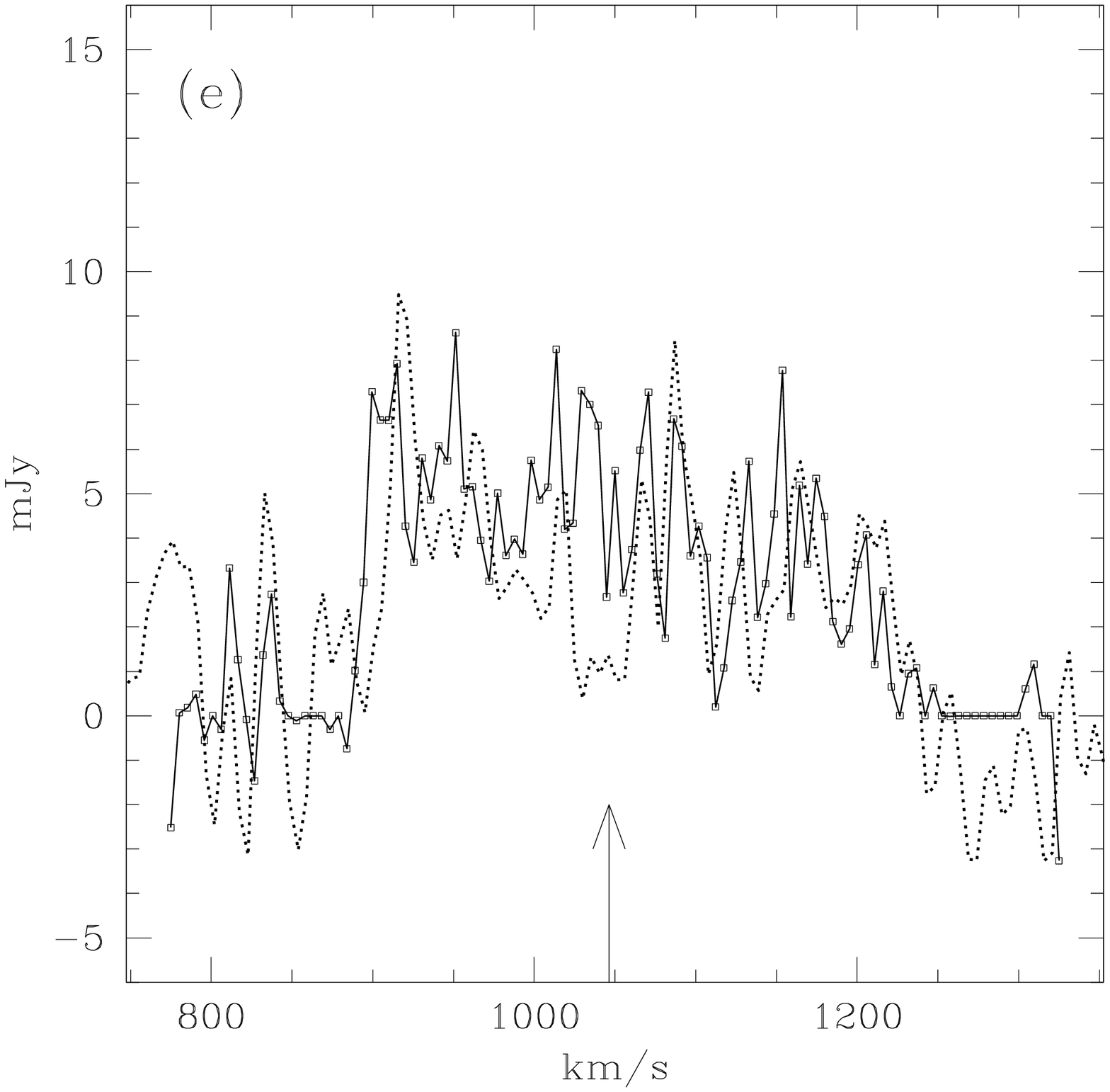} &\\

\end{tabular}
\figcaption{Continued.}
\end{figure}

\clearpage

 \begin{figure}
\figurenum{4}
 \begin{tabular}{cc}
 \includegraphics[width=0.45\textwidth, height=0.45\textwidth]{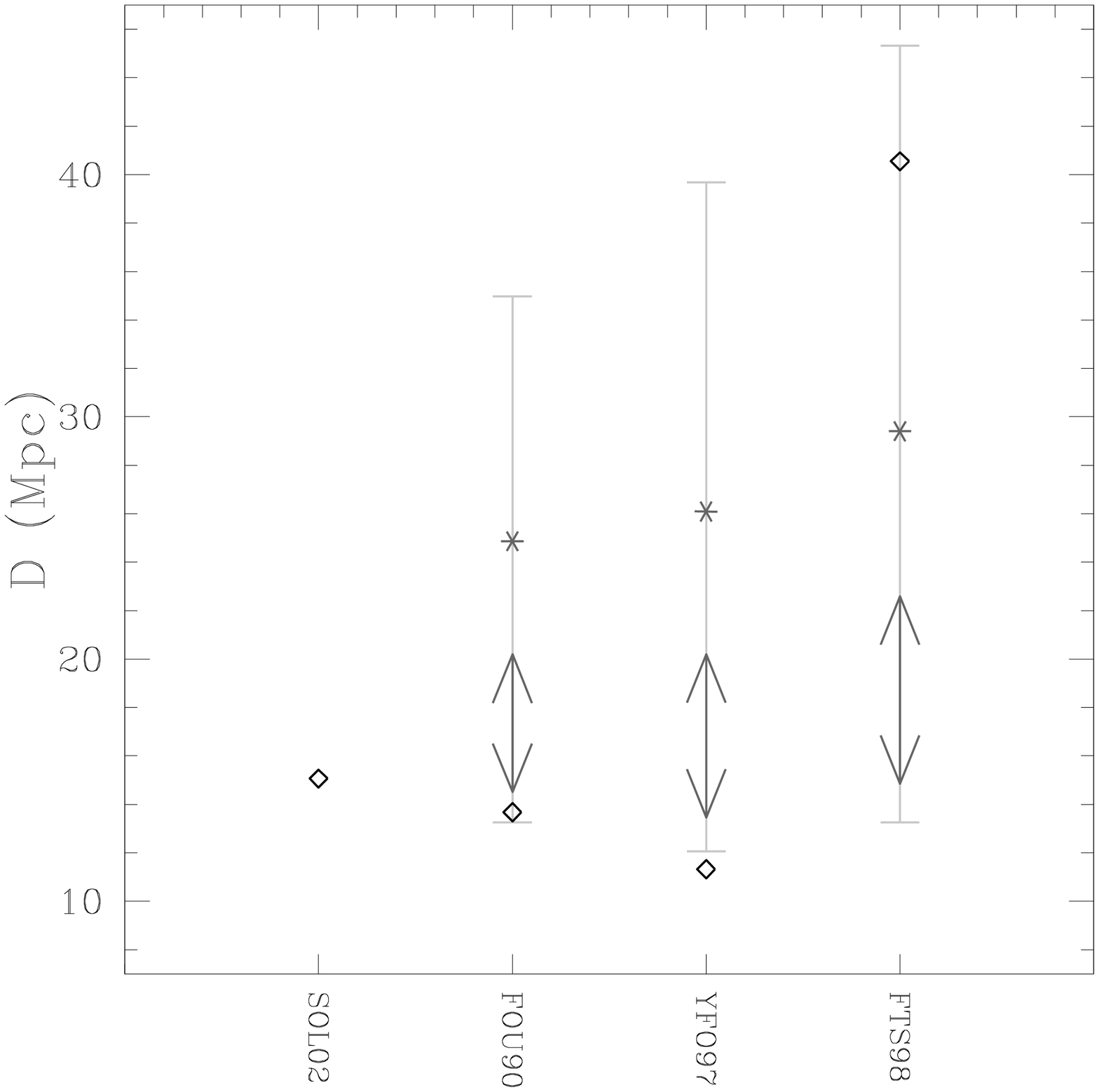}&
 \includegraphics[width=0.45\textwidth, height=0.45\textwidth]{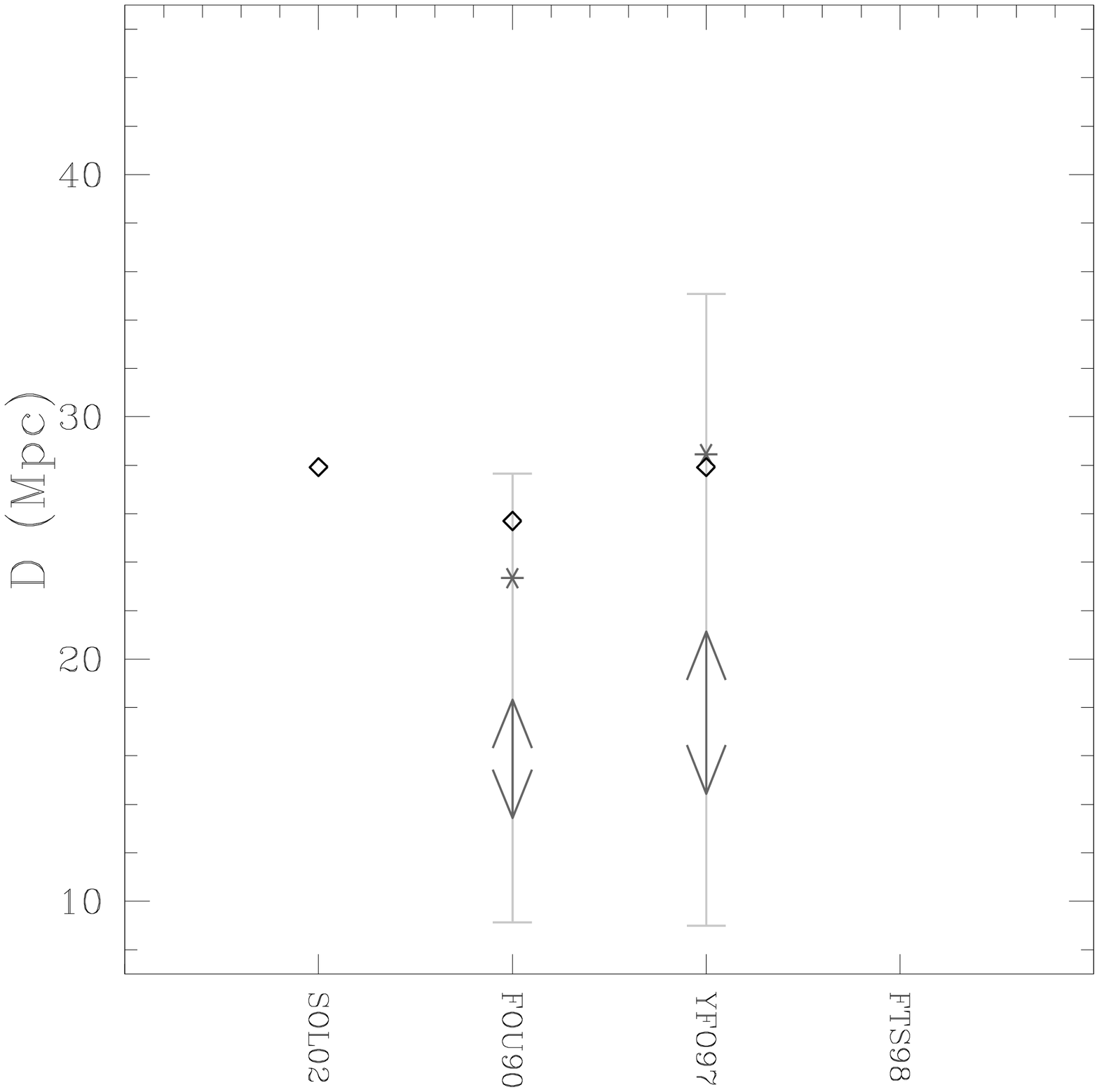}\\
 \end{tabular}
 \figcaption{TF distances to the disk galaxies NGC 4411A (a) and NGC
   4411B (b). The open diamonds indicate the radial distance to the
   galaxies published in several TF studies used in \citeauthor{Sol02},
   quoted in the horizontal axis (see caption of
   Fig.~\ref{tf_templates}). Vertical bars show the ranges of distances
   allowed by these studies when our VLA \hi\ linewidth estimates and
   the most extreme values of the inclination ever inferred for each
   object are used. Double pointed arrows encompass the ranges of
   distances corresponding to the values of inclinations and errors
   derived from our \hi\ data. Distance estimates resulting from the
   structural decomposition of the SDSS r-band images are indicated by
   asterisks.\label{tf_distances}}
 \end{figure}

 \begin{figure}
\figurenum{5}
 \includegraphics[width=10cm,angle=270]{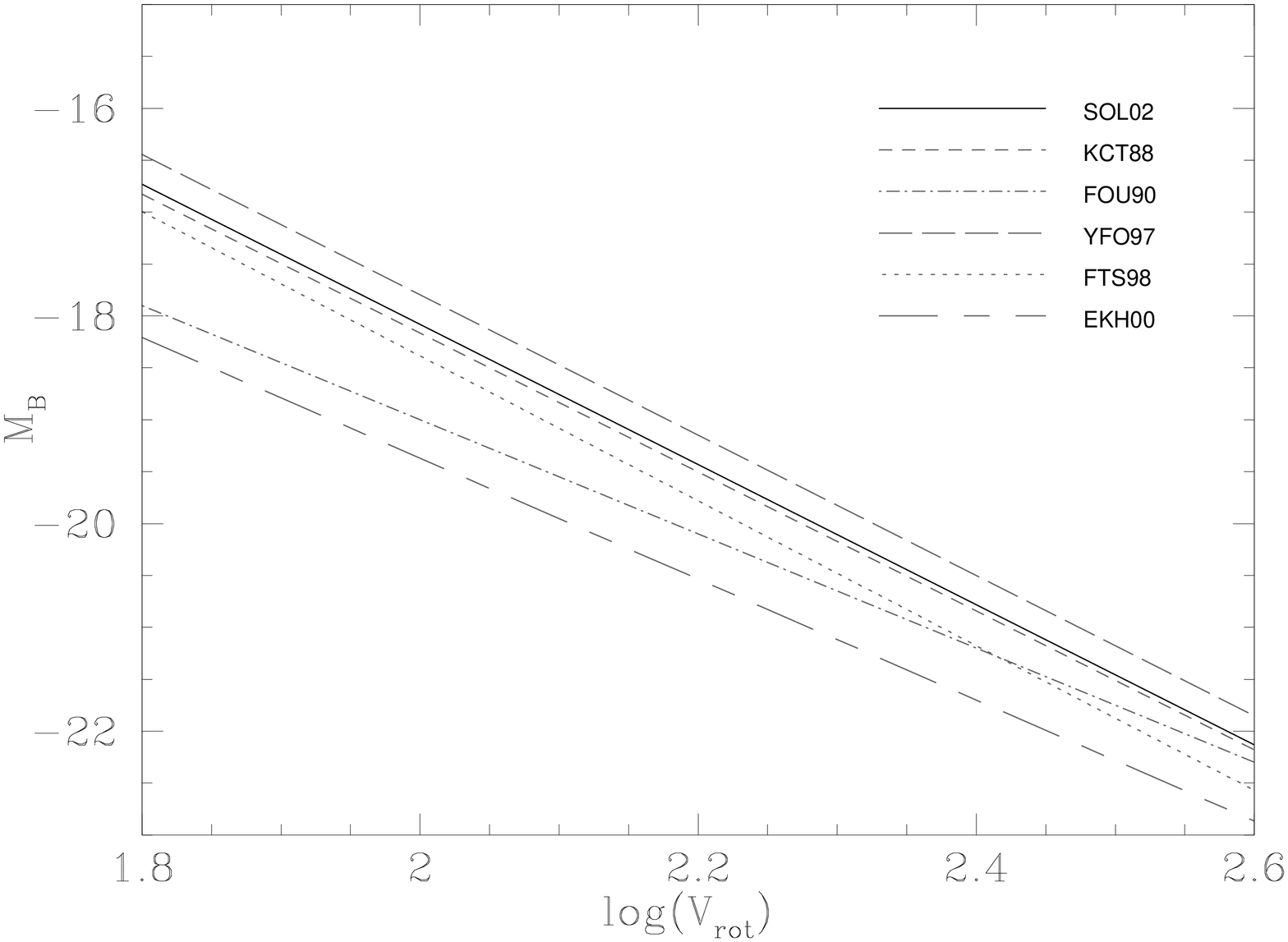} 
 \figcaption{$B$-band TF templates considered in
   \citeauthor{Sol02}. The acronyms correspond to \citet*{YFO97}
   (\citeauthor{YFO97}), \citet*{KCT88} (\citeauthor{KCT88}),
   \citet*{Fou90} (\citeauthor{Fou90}), \citet*{FTS98}
   (\citeauthor{FTS98}), and \citet*{Ekh00} (\citeauthor{Ekh00}). The
   relation resulting from the homogenization of the eight TF catalogs
   used by \citeauthor{Sol02} is also included. \label{tf_templates}}
 \end{figure}

%% file: toribio_tab1_selection.tex



\begin{deluxetable}{lcclrc@{\extracolsep{6pt}}c@{\extracolsep{6pt}}ccl}
\tabletypesize{\scriptsize}
\rotate
\tablecolumns{10}
\tablewidth{0pt}
\tablenum{1}
\tablecaption{Properties of the target galaxies\label{toribio_tab1}}
\tablehead{
\multicolumn{1}{c}{Galaxy} & \multicolumn{1}{c}{R.A.}  & 
\multicolumn{1}{c}{Dec.}  & \multicolumn{1}{c}{Type} & 
\multicolumn{1}{c}{$v_{\mathrm{hel}}$} &\multicolumn{1}{c}{ $a\times b$} 
& $i$ &
$m_{B}^{\mathrm{tot,c}}$ & DEF & \multicolumn{1}{c}{$d$}\\
\colhead{} & \multicolumn{2}{c}{(J2000)} & \colhead{} & 
\multicolumn{1}{c}{(km\ s$^{-1}$)} &
    \multicolumn{1}{c}{($\arcmin$)} & ($\degr$) & (mag) & \colhead{} & 
\multicolumn{1}{c}{(Mpc)}\\
    \multicolumn{1}{c}{(1)} & \multicolumn{1}{c}{(2)} & 
\multicolumn{1}{c}{(3)} & \multicolumn{1}{c}{(4)} & \multicolumn{1}{c}{(5)}
    & \multicolumn{1}{c}{(6)} & \multicolumn{1}{c}{(7)} & 
\multicolumn{1}{c}{(8)} & \multicolumn{1}{c}{(9)} & 
\multicolumn{1}{c}{(10)}
}
\startdata
\multicolumn{10}{l}{\hspace{-2mm}{\ \underline{Main targets}:}} \\
\noalign{\vspace{0.8mm}}
NGC 4307 &  12$^\mathrm{h}\!$22$^\mathrm{m}\!$05$\fs$6 & 
$+09\degr\!$02$\arcmin\!$37$\arcsec$ & Sbc & 1050 & $3.95\times 0.72$ & 
90.0 &11.83 &   1.24 & $27.4^{+3.5}_{-3.1}$\\

NGC 4356 & 12\phn  24\phn  14.4 & $+$08\phn 32\phn 09 & Sc& 1137 &  
$3.2 \times 0.5$ & 71.4 &13.24 &   1.35 & $29.4^{+3.7}_{-3.3}$ \\

NGC 4411B& 12\phn  26\phn  47.2 & $+$08\phn 53\phn 05 & Sc & 1271 &  
$3.45 \times 3.45$ &  26.7 &  12.81  & 0.63 & $27.9^{+0.7}_{-0.6}$\\

NGC 4492& 12\phn  30\phn  59.8 & $+$08\phn 04\phn 41 & Sa & 1777 &  
$1.96\times 1.96$ & 30.0 & 13.02 &  0.78 & $27.9^{+8.7}_{-6.7}$\\
\noalign{\vspace{0.8mm}}
\multicolumn{10}{l}{\hspace{-2mm}{\ \underline{Secondary targets}:}} \\
\noalign{\vspace{0.8mm}}
VCC 0740\tablenotemark{a} & 12\phn  24\phn  39.9 & $+$08\phn 30\phn 05 & 
Sm &  875 &  $0.71\times 0.32$ & 69.1 &15.27 & \nodata & 
\multicolumn{1}{c}{\nodata}\\
NGC 4411A\tablenotemark{b}& 12\phn  26\phn  30.1 & $+$08\phn 52\phn 18 & 
Sc  &  1282 & $2.79 \times 2.79$ & 54.4 &13.04 &  0.41 & 15.1 \\

VCC 0933\tablenotemark{b}& 12\phn  26\phn  44.3 & $+$08\phn 49\phn 25 & 
dE & 1381\tablenotemark{c} &  $\,\,0.5 \times 0.39$ &\nodata 
&17.00\tablenotemark{d}  & \nodata& \multicolumn{1}{c}{\nodata} \\

VCC 0976\tablenotemark{b}& 12\phn  27\phn  11.9 & $+$08\phn 50\phn 24 & 
dE & 1259\tablenotemark{c} &  $0.28 \times 0.18$ & \nodata & 
18.26\tablenotemark{d}  & \nodata & \multicolumn{1}{c}{\nodata}\\
\enddata
\tablenotetext{a}{: in the same field of NGC 4356.}
\tablenotetext{b}{: in the same field of NGC 4411B.}
\tablenotetext{c}{: derived from optical measurements as listed in the 
Galaxy On-Line Database Milano Network (GOLDMine).}
\tablenotetext{d}{: photographic magnitude.}
\tablecomments{Description of columns: \\
{\it (1)} NGC or VCC designations. \\  
{\it (2)}--{\it (3)} J2000 position listed
in the Lyon-Meudon Extragalactic Database (LEDA). \\
{\it (4)} Morphological type from GOLDMine.\\
{\it (5)} Heliocentric radial velocity obtained from \smallhi\ profile 
measurements.\\
{\it (6)} Angular size of the semi-major
 and semi-minor optical axes determined at the surface brightness of 25 
mag arcsec$^{-2}$ from GOLDMine.\\
{\it (7)} Inclination between the line-of-sight (LOS) and polar axis of 
the optical image listed in the LEDA.\\
{\it (8)} Total apparent $B$ magnitude corrected from inclination and 
extinction effects as listed in the LEDA.\\
{\it (9)} \smallhi\ deficiency parameter calculated using the definition 
in equation~(\ref{def}) and the optical sizes and \smallhi\
fluxes from conventional single-dish measurements attributed in the 
Arecibo General Catalog, a private compilation of 21 cm line
measurements maintained by R.\ Giovanelli and M.~P.\ Haynes at Cornell 
University. \\
{\it (10)} TF radial distance estimate from \citeauthor{Sol02}.
}
\end{deluxetable}


%% file: toribio_tab2_observations.tex
\begin{deluxetable}{l|c|c|c|c}
\tabletypesize{\scriptsize}

\tablecolumns{5}
\tablewidth{0pt}
\tablenum{2}
\tablecaption{Summary of the observations\label{toribio_tab2}}
\tablehead{ Field & \colhead{NGC 4307} & \colhead{NGC 4356} & 
\colhead{NGC 4492} & \colhead{NGC 4411B}}

\startdata

Array configuration&CS&CS&CS&CS\\
Observing dates & 09-07-2005& 16-07-2005& 12-08-2005 and & 01-10-2005 and\\
& && 18-09-2005& 02-10-2005\\
Total time on source (hr) &6.23&6.12&6.25&6.3\\
Phase center, $\alpha$ (J2000) & 12$^\mathrm{h}$ 22$^\mathrm{m}$ $05\fs 
6$ & 12$^\mathrm{h}$ 24$^\mathrm{m}$ $14\fs 5$ & 12$^\mathrm{h}$ 
30$^\mathrm{m}$ $47\fs 7$ & 12$^\mathrm{h}$ 26$^\mathrm{m}$ $47\fs 2$\\
Phase center, $\delta$ (J2000) & 09\degr\ 02\arcmin\ 37\arcsec & 
08\degr\ 32\arcmin\ 09\arcsec&  08\degr\ 04\arcmin\ 41\arcsec& 08\degr\ 
53\arcmin\ 04\arcsec \\

Flux calibrator (J2000)&1331+305&1331+305&1331+305&1331+305\\
Phase calibrator (J2000) & 1229+020& 1229+020& 1229+020& 1254+116\\
Bandwidth (1st/2nd IF pair, MHz)& 1.562/1.562 & 1.562/1.562 & 
1.562/3.125& 1.562/3.125 \\
Number of channels (1st/2nd IF pair) & 63/63 & 63/63 & 63/31& 63/31\\
Channel width\tablenotemark{a} (1st/2nd IF pair, kHz)& 24.4/24.4& 
24.4/24.4& 24.4/97.6& 24.4/97.6\\
Central heliocentric velocity\tablenotemark{b} &&&&\\
(1st/2nd IF pair,\kms)& 920/1180 &  940/1180& 1777/1777 & 1309/1309\\

\enddata

\tablenotetext{a}{: after Hanning smoothing.}
\tablenotetext{b}{: optical definition.}

\end{deluxetable}

%% file: toribio_tab3_deconvolution.tex
\begin{deluxetable}{l|c|c|c|c}

\tabletypesize{\scriptsize}
\tablecolumns{5}
\tablewidth{0pt}
\tablenum{3}
\tablecaption{Characteristics of the deconvolved images\label{toribio_tab3}}
\tablehead{Field
&\colhead{NGC 4307} & \colhead{NGC 4356} & \colhead{NGC 4492\tablenotemark{b}} & 
\colhead{NGC 4411B}}

\startdata
Robustness parameter &0.7&0.7&1&0.7\\
Gaussian taper width at 30\% level (k$\lambda$) &0  &0&9&0\\
Synthesized beam FWHM ($\arcsec$) & 16.09 $\times$ 15.26 & 16.29  
$\times$ 15.47 & 23.03  $\times$ 21.18 & 16.74  $\times$ 14.11\\
Synthesized beam position angle ($\degr$) & $-51.1$& $-55.7$& $-55.5$& 
$-38.8$\\
rms noise per channel in line-free channels (mJy beam$^{-1}$) & $\sim 
0.57$\tablenotemark{a}&$\sim 0.60$\tablenotemark{a}&
0.35 & $\sim 0.60$ \\
1$\sigma$ limiting column density per channel ( $10^{19}$ cm$^{-2}$)&
 1.32& 1.36 & 1.65 &1.46\\
rms noise in total intensity map (mJy beam$^{-1}$\kms)& 10.3& 11.1& 
16.9 &9.0\\
1$\sigma$ limiting column density in total intensity map &&&&\\
( $10^{19}$  cm$^{-2}$\kms) &4.6&4.8&3.7&4.2\\
\enddata

\tablenotetext{a}{: $\sim 0.42$ for the cube obtained by averaging each 
two channels.}
\tablenotetext{b}{: corresponding to images obtained from IF2 (see Table~\ref{toribio_tab3}).}

\end{deluxetable}

%% file: toribio_tab4_analysis.tex
\begin{deluxetable}{lrrrrrr|rrrr}
\rotate
\tabletypesize{\scriptsize}
\tablecolumns{10}
\tablewidth{0pt}
\tablenum{4}
\tablecaption{Results from our VLA data and recent Arecibo 
measurements\label{toribio_tab4}}
\tablehead{

  \colhead{}& \multicolumn{6}{c}{VLA} &
  \multicolumn{4}{|c}{ALFALFA}  \\

\colhead{Galaxy}& \colhead{$\wtwenty$} & \colhead{$\wfifty$} & 
\colhead{$\int{S\;dv}$} &\colhead{$\df$}&\colhead{$v_\mathrm{sys}$}  
&\colhead{$D\subhi$}& \multicolumn{1}{|c}{$\wfifty$} &  
\colhead{$\int{S\;dv}$} &\colhead{$\df$}&\colhead{$v_\mathrm{sys}$}\\
& \colhead{(km\ s$^{-1}$)} & \colhead{(km\ s$^{-1}$)} & 
\colhead{(Jy\kms)} &
\colhead{}& \colhead{(km\ s$^{-1}$)}& \colhead{($\arcmin$)} &   
\multicolumn{1}{|c}{(km\ s$^{-1}$)} &
 \colhead{(Jy\kms)} &  \colhead{}&  \colhead{(km\ s$^{-1}$)}  \\
\colhead{(1)}&\colhead{(2)}&\colhead{(3)}&\colhead{(4)}&\colhead{(5)}&\colhead{(6)}&\colhead{(7)}&
\multicolumn{1}{|c}{(8)}& \colhead{(9)}& \colhead{(10)}& \colhead{(11)}

}
\startdata

NGC 4307&  327$\pm$ 8 & 286$\pm$15 &  1.48$\pm$ 0.09& 1.32  & 
1048$\pm$4 & 1.83$\pm$0.16& 314$\pm$41 &1.21$\pm$0.10 & 1.41 & 
1065$\pm$21 \\
NGC 4356 & 204$\pm$ 7 & 154$\pm$27 & 0.44$\pm$ 0.04&  1.71 & 1120$\pm$6
& 1.41$\pm$0.14 &237$\pm$25  & 0.87$\pm$0.08& 1.41 & 1092$\pm$12 \\
NGC 4411B & 93$\pm$ 1& 77$\pm$ 1 &18.91$\pm$ 0.95 & 0.13 &1270$\pm$0 & 
4.03$\pm$0.29&82$\pm$ 2 &17.57$\pm$0.07 & 0.16    & 1272$\pm$ 1 \\
NGC 4492\tablenotemark{a} & 141$\pm$12& 79$\pm$25& 0.43$\pm$ 0.04 & 1.21 &1768$\pm$7 & 
1.38$\pm$0.17 & 182$\pm$ 9  &0.59$\pm$0.07 & 1.07& 1740$\pm$ 5 \\
VCC 740\tablenotemark{b} & 132$\pm$ 3 & 109$\pm$ 5 &  1.54$\pm$ 0.09  & 
$-0.21$ & 875$\pm$1  & 1.47$\pm$0.14 & 97$\pm$ 5&  1.39$\pm$0.06 & 
$-0.17$ & 877$\pm$ 2  \\ 
NGC 4411A\tablenotemark{c} & 107$\pm$ 1& 90$\pm$ 1 & 12.47$\pm$ 0.63  & 
$-0.03$ &1278$\pm$0&  3.84$\pm$0.27 & 89$\pm$ 0& 13.6$\pm$0.07 & 
$-0.07$ & 1278$\pm$ 0\\

\enddata
\tablenotetext{a}{: Columns (2)--(6) show measurements from IF1 integrated
profile (resolution $\sim$ 5.2 \kms), whereas (7) was measured on IF2 images.}
\tablenotetext{b}{: in the same field of NGC 4356.}
\tablenotetext{c}{: in the same field of NGC 4411B.}

\tablecomments{Column description:\\
{\it (1)} Target name.\\
{\it (2)}--{\it (3)} Width of the \smallhi\ line profile measured at 
20$\%$ and 50$\%$ of the peak flux, respectively. Adopted broadening 
corrections correspond to a spectral resolution of 5.2\kms\ (see text). \\
{\it (4)} Integrated total flux observed with the VLA.\\
{\it (5)} Corresponding value of the \smallhi\
deficiency parameter recalculated from our VLA measurements.\\
{\it (6)} Heliocentric systemic velocity.\\
{\it (7)} \smallhi\ diameter measured at a column density of
$10^{20}$ atoms~cm$^{-2}$ (no correction for beam smearing is applied).\\
{\it (8)} Width of the \smallhi\ line profile at 50$\%$ of the peak flux 
measured by ALFALFA.\\
{\it (9)} Integrated total flux measured by ALFALFA.\\
{\it (10)} Corresponding value of the \smallhi\
deficiency parameter recalculated from ALFALFA measurements.\\
{\it (11)} Heliocentric systemic velocity measured by ALFALFA.\\
}

\end{deluxetable}